# Analysis and simulation of the operation of a Kelvin probe.

Robert D. Reasenberg,[1,] Kathleen P. Donahue,[1,2] and James D. Phillips [1]

[1] Smithsonian Astrophysical Observatory, Harvard-Smithsonian Center for Astrophysics, 60 Garden St., Cambridge, MA 02138.

[2] Current address: Department of Physics, Harvard University, Cambridge, MA 02138.

E-mail: reasenberg@cfa.harvard.edu

**Abstract**. Experiments that measure extremely small gravitational forces are often hampered by the presence of non-gravitational forces that can neither be calculated nor separately measured. Among these spurious forces is electrostatic attraction between a test mass and its surroundings due to the presence of spatially varying surface potential known as the "patch effect." In order to make surfaces with small surface potential variation, it is necessary to be able to measure it. A Kelvin probe (KP) measures contact potential difference (CPD), using the time-varying capacitance between the sample and a vibrating tip that is biased with a backing potential. Assuming that the tip remains constant, this measures the sample's surface potential variation. We examine the operation of the KP from the perspective of parameter estimation in the presence of noise. We show that, when the CPD is estimated from measurements at two separate backing potentials, the standard deviation of the optimal estimate depends on the total observing time. Further, the observing time may be unevenly divided between the two backing potentials, provided the values of those potentials are correspondingly set. We simulate a two-stage KP data analysis, including a sub-optimal estimator with advantages for real-time operation. Based on the real-time version, we present a novel approach to stabilizing the average distance of the tip from the sample. We also present the results of a series of covariance analyses that validate and bound the applicability of the suboptimal estimator, make a comparison with the results of an optimal estimator and guide the user. We discuss the application of the KP to the LISA and to a test of the weak equivalence principle.

Keywords: equivalence principle, WEP, Kelvin probe, surface potential, CPD, patch effect.

PACS numbers: 04.80.−y, 06.30.−k, 07.07.Df, 07.50.Ek, 73.40.Cg

## 1. Introduction

The sounding-rocket based principle of equivalence measurement (SR-POEM) [Reasenberg et al. 2012] will determine η, the coefficient of equivalence principle violation, with an uncertainty of $2 \times 10^{-17}$ from a single flight. This will provide a four order of magnitude advance over the current best test [Schlamminger 2008, Wagner 2012] and two orders advance over the MICROSCOPE Mission which is now predicted to fly in 2016 [Touboul 2012]. An important spurious force in the SR-POEM experiment is electrostatic attraction between the test masses and their surroundings due to the presence of spatially varying surface potential known as the "patch effect." In order to make surfaces with small surface potential variation, it is necessary to be able to measure the small potential. This paper describes one aspect of our program to make a surface with a small enough electrostatic (patch effect) contribution to the SR-POEM error.

A Kelvin probe (KP) is a tool for measuring contact potential difference (CPD, designated $V_c$ ), and thus both the spatial and temporal variation of surface potential. It has a vibrating capacitance plate (tip) that is placed near a conducting surface of interest. When an electrical connection is made between the two electrodes, their Fermi levels equalize producing a potential difference between the opposing plates. The vibration, which causes a changing capacitance, results in a current that can be measured. If a backing (bias) potential ($V_b$) is added between the tip and the surface, the current is proportional to $V_b$-$V_c$. From the relationship between the measured current and the (adjustable) backing potential, one can determine the CPD between the tip and the surface under study. The magnitude of the measured current depends on the mean spacing between the tip and the surface, among other things. Therefore, this distance needs to be measured independently. Any change found can be fed back to stabilize the distance.

The KP is an important surface science and engineering tool [Baikie et al. 1991, Subramanyam 2009]. Past fundamental physics experiments requiring uniform surface potential include searches for fractional electric charges on stable matter [Phillips et al. 1988], tests of the force of gravity on electrons and positrons [Lockhart 1977, Henderson 1987, Bardeen 1988] and on antiprotons [Camp 1991, 1992, 1993], and the Gravity Probe-B test of the dragging of inertial reference frames [Buchman 2011]. Continuing experiments requiring uniform surface potential include measurements of the Casimir force (see below), the planned LISA (Laser Interferometer Space Antenna) space-based gravitational wave observatory [Pollack 2008], and tests of the equivalence principle [Reasenberg 2012, Touboul 2012]. The LIGO (Laser Interferometer Gravitational Wave Observatory) group is concerned with issues of potential disturbances from the boundaries of their experimental volume, but only about disturbances at a relatively high frequency, on insulators [Harry 2008, D. Shoemaker, priv. comm. 1/26/2012].

Much has been written recently about the influence of the electrostatic force in experiments on the Casimir effect [Kim et al 2010]. Modern Casimir experiments are usually conducted in a sphere-plane geometry. At greater distance, the diameter of the region in which the surface potential contributes importantly to the electrostatic energy increases. Also, greater distance reduces the contribution of surface potential variations of high spatial frequency. Therefore, the electrostatic force varies with distance in a manner that depends on the spatial distribution of surface potential. Speake & Trenkel [2003] relate the force to the statistical properties of the surface potential distribution. The change of the Casimir force with distance is the quantity measured. While other systematic effects are also important, at present, separating the electrostatic force from Casimir measurments depends on assumptions about the statistics of the surface potential distribution. As a recent review stated [Lambrecht 2012], "A better characterization of the patches is a crucial condition to reaching firmer conclusions."

SR-POEM uses laser gauges to determine the relative acceleration of a pair of test masses in free fall inside a sounding-rocket payload. The test masses are surrounded by a set of capacitance plates used both for sensing motion in secondary degrees of freedom and for electrostatically forcing test-mass motion. The estimated uncertainty in $\eta$ is 2 $10^{-17}$ from a single flight that includes eight separate "drops" of 120 seconds each. Between drops, the entire payload is inverted using gas jets, thus reversing the direction of the potential equivalence principle violation in payload coordinates and cancelling many kinds of measurement bias. This experiment requires that spurious forces on the test masses be both small and stable. In particular, the random variation of surface potential on the test masses and the facing capacitance plates must not change from one drop to the next by more than 0.1 mV RMS. In order to reliably make such a surface, one must be able to measure it.

The spatial variation and, to a limited extent, the time variation of surface potential has been measured. Both Camp et al., [1991] and Robertson et al. [2006] achieved a resolution in potential of 1 mV, with spatial resolution of 1.5 and 3 mm, respectively. Several surfaces exhibited spatial variation of ~1 mV rms. Robertson measured stability of the potential averaged over a single circular patch of diameter 3 mm for several hours, finding a fluctuation of $1\ \text{mV}\ \text{Hz}^{-1/2}$ from $4 \times 10^{-4}$ to $2 \times 10^{-1}$ Hz.

Popescu [2011] has investigated the relative merits of flat (parallel plate) tips and spherical tips. He concluded that, when compared to flat tips, "spherical tips yield poorer performance in terms of both signal amplitude and parameter standard deviations over the entire range of $\gamma$ ." He further noted that "only the flat tip models … allow for the full control of the instrument based on the estimated values of the parameters." Based on his published and our unpublished analyses, we restrict our attention to flat-tipped instruments.

The KP tip and the sample form a nearly parallel plate capacitor with area $\mathcal{A}$ (which is typically circular) and separation $d$, where we assume that $d$ is much smaller than the minimum lateral dimension of the capacitor plate. For the present analysis, we ignore both fringing fields and the tilt of the KP tip with respect to the sample. For a mean tip-sample distance of $d_0$ and a sinusoidal vibration of amplitude $d_1$, the time-varying spacing is

$$d = d_0(1 - \gamma\, cos(\varphi)) \tag{1}$$

where $\gamma \equiv d_1/d_0$, $\varphi \equiv 2\pi f t$, and $f$ is the vibration frequency. Over the expected range of d, the principal effect of the fringing field is to add a fixed capacitance in parallel with the variable capacitor

The time varying capacitance is therefore

$$C = \frac{\epsilon_0 \mathcal{A}}{d_0(1 - \gamma\, cos(\varphi))} \tag{2}$$

where $\epsilon_0$ is the permittivity of vacuum (~8.854 pF/m). If there is a potential of $V_g = V_b - V_c$ across the gap, the instantaneous charge will be $CV_g$ and the time variation will result in a current

$$i = \frac{-\gamma \epsilon_0 \mathcal{A}\, V_g\, \omega \sin(\varphi)}{d_0\big(1 - \gamma\, cos(\varphi)\big)^2} \tag{3}$$

where $\omega = 2\pi f$. That current enters an amplifier whose input has a low impedance and whose output is $iG$ (in Volts), where G is the frequency-independent transimpedance gain (in Ohms). This paper addresses the analysis of the signal after it has been amplified and digitized.

The parameters of a KP that appear in Eq. 3 are $d_0$, $d_1$, $\mathcal{A}$, $V_b$, and $\omega$. Of these, $\mathcal{A}$, $V_b$, and $\omega$ do not vary importantly. However, the tip-sample distance $d_0$, typically 100 $\mu m$, is subject to significant variation due to thermal expansion and elastic response in the equipment holding the vibrating tip near the sample. This is particularly so when the tip is scanned over the surface to measure spatial variation. $d_0$ is usually maintained by servo control. Often, the signal used to sense $d_0$ is the "gradient," which is the marginal ratio of the signal to the backing potential. In one study of the spatial distribution of surface potential, a striped pattern was observed in the maps of gradient [Robertson et al. 2006]. Although in that case the gradient was under servo control, it changed by 4%, depending on whether the probe was scanning in the $x$ or $-x$ direction. Since the probe-sample distance, $d_0$, in this work was $\sim$100 $\mu m$, the distance change was $\sim$4 $\mu m$. A distance change of this sort alters the signal strength. If the backing potentials used are not symmetrically distributed around $V_c$, i.e., if the KP is not operated in the "standard KP operating mode" described below, the distance changes introduce a bias in the estimate of $V_c$. In fact, in some of their $V_c$ maps, Robertson et al. did point out that there were clear traces of a pattern similar to the striped pattern in the maps of the gradient. The peak-to-peak amplitude was $\sim$1 $mV$. (We are not sure what mechanism produced this result.)

Changes in the probe oscillation amplitude $d_1$ are generally less significant. For almost all studies, particularly those in support of SR-POEM, stability for periods of an hour or less will suffice. Over this time, the temperature of a laboratory is unlikely to change by more than 1°C. If the oscillating tip is suspended from flexure springs and driven magnetically, this temperature change is unlikely to produce a change $\Delta d_1/d_1$ more than a tenth of the change $\Delta d_0/d_0$ described above, so the stability of $d_0$ is the more important issue.

In the following sections, we investigate the operation and data analysis for an ideal Kelvin probe. In Section II, we find that for each spot on the surface of a sample, a minimum variance estimate of CPD at a given epoch can be obtained by using two values of backing potential equally spaced above and below the CPD and observing for equal times at these backing potentials. The description is generalized to yield a single-parameter family of observing patterns having two backing potentials, the same total observing time and the same variance. In Section III, we investigate the sensitivity of a Kelvin probe and find that, with a modern preamplifier, it can reach a precision of a few µV in one second of measuring. (Systematic error, as from the triboelectric effect or vibration at the probe oscillation frequency, will be addressed in a future paper.) A series of simulations is discussed in Section IV, where we also consider real-time estimation of CPD, real-time estimation of $d_g$, the distance of closest approach of the vibrating capacitor plate and the sample, and the measurement of a rapidly varying CPD. In Section V, we discuss using the real-time estimates to obtain a suboptimal estimate of CPD that should be adequate for many scientific and engineering purposes. Finally, we offer concluding remarks in Section VI.

## 2. Strategy for Setting $V_b$

To measure the CPD at a single spot on the surface under study, the vibrating probe tip is placed over the spot and current measurements are made at one or more values of the backing potential, $V_b$. Here we find an optimal strategy for selecting the number of values of $V_b$ in the sense of yielding the lowest uncertainty in the estimate of $V_c$ in a fixed elapsed time. (Measuring time, $\tau_0$, is smaller due to overhead such as waiting for a new value of $V_b$ to settle.) We start with the assumption that three values of $V_b$ will be useful. We then show that equally good CPD estimates result from having two or even one value of $V_b$. However, using a single $V_b$ complicates the job of regulating $d_0$. Next we look at having a larger number of $V_b$ by combining data from two or more sets of measurements that each make optimal use of two values of $V_b$. From this analysis we show that the optimal number of values of $V_b$ is two.

We take the initial backing potentials as

$$V_{bi} = \begin{cases} V_1 + V_0, & i = 1 \\ V_1 - V_0, & i = 2 \\ V_2, & i = 3 \end{cases} \tag{4}$$

where the measurement at $V_{bi}$ is made for time $\tau_i,\ i = 1, 2, 3$. We assume the measurements are corrupted by white noise with fixed density and we perform a covariance analysis to determine what values $V_0$, $V_1$, and $V_2$ should take.

We divide the analysis into Stages 1 and 2. A Stage 1 analysis results in $\hat{h}_i$, estimates of the peak amplitude of the amplified and digitized capacitor current. For each of the backing potentials, $V_{bi}$, $\hat{h}_i$ must be found by fitting to some or all of the KP data taken at $V_b = V_{bi}$, as discussed in later sections. We introduce a simple model of $h_i$

$$h_i = \alpha(V_{bi} - V_c), \quad i = 1, 2, 3 \tag{5}$$

A Stage 2 analysis follows a Stage 1 analysis and produces an estimate of CPD. We form the coefficient matrix of the weighted least-squares (WLS) estimator with the parameter set $\{\alpha, V_c\}$ and weights $W_0\, \tau_i$ (see Section III), where $\tau_i$ is the observing time on which $\hat{h}_i$ is based. We assume that the $\hat{h}_i$ are uncorrelated and based on the analysis of KP data accumulated in a total time $\tau_0 = \tau_1 + \tau_2 + \tau_3$, but do not restrain the separate times. Inverting the coefficient matrix and taking the square root of the diagonal elements gives us the parameter uncertainties as functions of $V_0$, $V_1$, $V_2$, $V_c$, $\tau_1$, $\tau_2$, and $\tau_3$. In particular, for the CPD:

$$\sigma(V_c) = \sqrt{\frac{4\tau_1 V_0(V_c - V_1) - \tau_0(V_c - V_1)^2 + 2(\tau_0 - \tau_3)V_0(V_1 - V_c) + \tau_3(V_1 - V_2)(V_1 + V_2 - 2V_c) + (\tau_3 - \tau_0)V_0^2}{\alpha^2 W_0\,(4\tau_1^2 V_0^2 - 4\tau_1 V_0((\tau_0 - \tau_3)V_0 + \tau_3(V_1 - V_2)) + \tau_3(\tau_3 - \tau_0)(V_0 - V_1 + V_2)^2)}} \tag{6}$$

For the interesting special case of $\tau_3$=0, this becomes

$$\sigma(V_c)|_{\tau_3=0} = \frac{1}{2\alpha V_0}\sqrt{\frac{\tau_0(V_0 + V_c - V_1)^2 - 4\tau_1 V_0(V_c - V_1)}{W_0(\tau_0 - \tau_1)\tau_1}} \tag{7}$$

Considering Eq 6, if we minimize $\sigma(V_c)$ with respect to $V_1$, we get

$$V_1 = \frac{(\tau_0 - 2\tau_1 - \tau_3)V_0 + \tau_0 V_c - \tau_3 V_2}{\tau_0 - \tau_3} \tag{8}$$

and, applying Eq 8 to Eq 6, we find

$$\sigma(V_c) = \frac{1}{\alpha\sqrt{\tau_0 W_0}} \tag{9}$$

Note that this is independent of $V_0$ and $\tau_3$. If we minimize Eq 6 with respect to $\tau_1$, we get the same result.

Since Eq 9 is independent of $\tau_3$, there is no advantage to having a non-zero $\tau_3$. However, there is a disadvantage: each time $V_b$ is changed, there is a required settling time (i.e., non-productive time) before data can again be taken. We will therefore assume $\tau_3 = 0$ (i.e., that there are only two backing potentials used) for all of the following analyses. Further, if all of the time is spent at the first potential, i.e., $\tau_1 = \tau_0$, then Eq 4 and 8 yield $V_b = V_c$. This is an operating mode usually associated with a servo that adjusts $V_b$ to make it equal to $V_c$. The servo-based approach has two disadvantages: the servo needs to settle before useful data can be taken, and $d_0$ must be sensed independently

There is another mode in which a single value of $V_b$ is used. In order to regulate $d_0$, a value of $V_b$ different from $V_c$ would be employed, so that the signal current is non-zero. Then a measure of the functional form of the current as a function of time in Eq 3 such as the ratio of the signal at frequency $2\omega$ to that at $\omega$ yields an estimate of $\gamma = d_1/d_0$. Assuming that $d_1$ has remained constant, we can regulate $\gamma$ so as to keep $d_0$ constant. In this mode, changes in $V_b$ are not needed, so time does not need to be spent settling. Occasional tests are made, changing $V_b$, to guard against changes in $d_1$ or the amplifier gain.

A simple special case of the solution defined by Eq 8 with $\tau_3 = 0$, which we will call the "standard KP operating mode," is $V_1 = V_c$ and $\tau_1 = \tau_2 = \tau_0/2$. The requirement that $V_1 = V_c$ has previously been found assuming the observing times are equal [Reasenberg 2010, Popescu 2011]. We note that the correlation, $\rho(\alpha, V_c)$, (from the WLS covariance; $\alpha$ is defined in Eq. 5) is proportional to $(V_c - V_1)/V_0$ and so vanishes in the standard KP operating mode. From Eq 7, we can find the impact of an error in the nominally equal division of time or equality of spacing of $V_b$ around $V_c$. In particular, $\sigma(V_c)|_{\tau_3=0}$ will be within 10% of its minimum value if $V_1$=$V_c$ and $0.3\ \tau_0 < \tau_1 < 0.7\ \tau_0$ or if $\tau_1 = \tau_2 = 0.5\ \tau_0$ and $|V_1$-$V_c| < 0.45\ V_0$.

Finally, there is one additional operating mode worthy of mention. It is characterized by $V_1 = 0$ and thus $V_{b1} = -V_{b2} = V_0$, which removes the need for a pair of precision adjustable power supplies for the backing potentials. The optimal estimate with error given by Eq 9, is achieved by using time division found from Eq 8: $\tau_1 = \tau_0\ (V_0 + V_c)/(2\ V_0)$. It will be seen later that the estimate of $d_0$ depends on $V_b$-$V_c$. Making $V_b$ symmetric around zero instead of around $V_c$ degrades the estimate of changes in $d_0$. However, as long as $V_0 \gtrsim 2\ V_c$, this is unlikely to matter.

Equation 9 tells us that any observation sequence with a given $\tau_0$ (e.g., with $\tau_3 = 0$, and two values of $V_b$) will give the same accuracy once the backing potentials or division of time is optimized. Consider a pair of such observation sequences, A and B, for a single spot with $V_1 = V_c$. Assume independent measurement times, $\tau_{0A}$ and $\tau_{0B}$, and different backing potentials, i.e., {$V_{bA1,}$ $V_{bA2}$ } different from {$V_{bB1,}$ $V_{bB2}$ }. Then, since $\rho(\alpha, V_c) = 0$, if we combine the data from sequences A and B, we get the same result, with the same uncertainty, as if we took the weighted average of the two separate results. Thus, the uncertainty found for the combined data set is given by Eq 9 with $\tau_0$ replaced by $\tau_{0A} + \tau_{0B}$. In this way, we can analyze a complex observing sequence with many values of $V_0$, each with a separate observing time. Provided that all sequences were taken in the standard KP operating mode, the uncertainty in the estimate of $V_c$ will depend only on the total observing time. Since each change of $V_b$ uses some time for settling, the best result for a given elapsed time (observation plus settling time) will use just one value of $V_0$, and $\tau_3 = 0$. This includes the standard KP operating mode: $\tau_1 = \tau_2 = \tau_0/2$.

## 3. Data Models and Analyses

In this section, we look at the Stage 1 analysis of the digitized electronic signal produced by the KP hardware. That analysis can yield a "single cycle solution," from which we might get the real-time estimates of $d_0$, and the $h_i$ discussed in Section II. We then consider two alternative Stage 2 analyses.

We consider a vibrating KP capacitance plate (tip) with linear dimensions large compared to the mean distance to the sample under study. For example, consider a disk of 1 mm radius held at a mean distance, $d_0$ = 0.1 mm, from the sample and vibrating at 100 Hz. For a minimum tip-sample separation of 0.01 mm, $\gamma = 0.9$. We assume the first electronic stage is an amplifier with zero input impedance whose output voltage is proportional to its input current. We further assume that its gain is substantially independent of frequency over the frequency range from $f$ to $\beta f$, where $\beta$ is large enough to incorporate all harmonics containing significant information as discussed below. Then the current generated by the vibrating plate is converted to a potential $z(\varphi)$ given by

$$\begin{aligned} z(\varphi) &= Y\, F(\gamma,\varphi) \\ Y &= \frac{\omega\epsilon_0 \mathcal{A} G}{d_0}\,(V_b - V_c) \\ F(\gamma,\varphi) &= \frac{-\gamma\sin\varphi}{(1-\gamma\cos\varphi\,)^2} \end{aligned} \tag{10}$$

where $\varphi = \omega t$, $\mathcal{A}$ is the area of the vibrating KP capacitance plate, $G$ is the gain (digitized signal divided by input current), $V_b$ is the backing potential, and $V_c$ is the sought-after CPD between the sample and the KP tip. To avoid aliasing, the signal is limited to a maximum frequency $f_0$, and the sample rate $f_S \geq 2f_0$. In Eq 10, we have neglected "fringing fields," which would modify $F(\gamma,\varphi)$. Figure 1 shows $F(\gamma,\varphi)$ for a few values of $\gamma$. Based on the results of Section II we adopt the standard KP operating mode in which, for each point on the surface, we use two values of $V_b$ chosen so that their average is $V_c$. The observing time will be evenly split between these two.

We assume that the input current noise of the amplifier dominates the measurement error. In our present preamplifier design, this noise is primarily due to Johnson noise in the feedback resistor. We treat white noise with density $\sigma_j$ (in A $Hz^{-1/2}$ ). Then the noise at the output of the amplifier and therefore on the measurements of $z(\varphi)$ is $\sigma_z = G\,\sigma_j$ (in V $Hz^{-1/2}$ ). (We will address noise sources that precede the preamplifier, such as vibration at the probe oscillation frequency, in a future paper.) The weights $W_0\tau_i$ introduced in Section II become $\tau_j/\left(G\,\sigma_j\right)^2$.

*3.1. Estimating phase and height of peaks - real time case (Stage 1 analysis).*

We consider the analysis of a subset of the data during a single cycle of vibration: $-\pi < \varphi < \pi$. We seek to estimate $\varphi_0$ and $h$, where $\varphi_0$ is the absolute value of the phase at the positive and negative peaks and $h$ is the corresponding amplitude. To find $\varphi_0$ we start from Eq 10 and solve $dF/d\varphi = 0$ for $\cos(\varphi)$. This gives us $\varphi$ at the extreme of $F(\gamma, \varphi)$, where $|\varphi| \equiv \varphi_0$. From $\cos(\varphi_0)$ we get

$$\varphi_0 = \mathrm{acos}\left(\frac{\sqrt{1+8\,\gamma^2}-1}{2\,\gamma}\right) \tag{11}$$

and, for $0 \le \gamma \le 0.993$, the approximation $\varphi_0 \approx (\pi - 3\gamma)/2$ is within 20% of the value of $\varphi_0$.

Given an estimate of $\varphi_0$, one can find a corresponding estimate of $\gamma$

$$\gamma(\varphi_0) = \frac{\cos(\varphi_0)}{2-\cos^2\,(\varphi_0)} \tag{12}$$

From that, one gets $F_0 \equiv |F(\gamma(\varphi_0), \varphi_0)|$, the absolute value of $F$ at the peak, and $h = YF_0$.

If we assume that at least on short time scales the oscillation amplitude, $d_1$, is stable, as discussed in Section I, then position feedback could be based on the estimate of $\varphi_0$: $1/d_0 = \gamma(\varphi_0)/d_1$. Figure 2 shows that $1.1 < d(\varphi_0)/d\gamma < 2.0$. Thus, feedback on $d_0^{-1}$ can be based directly on the estimate of $\varphi_0$. During setup, a value for $d(\varphi_0)/d\gamma$ can be calculated and then used for the run. Since $d(\varphi_0)/d\gamma$ only appears in the loop gain, and only as a multiplicative factor, it is not required to be precisely determined or even very stable and it would suffice to estimate it from the parameters established during setup.

While the KP is in data-taking mode, $\varphi_0$ can be estimated separately during each cycle of the tip vibration. It is a separate matter to determine the correct response time of the position servo and the corresponding degree of averaging of the noisy, single-cycle estimates of $\varphi_0$. The design of that servo is not addressed in this paper, and depends on the disturbance to $d_0$.

For the real-time Stage 1 analysis, we could take all of the data from a single cycle of the vibration and fit the parameters of Eq 10. Alternatively, the data from N cycles could be accumulated, averaging all N data taken with the same phase, $\varphi_m, m = \{1 \ldots M\}$. This approach benefits from the data sampling rate being made an exact multiple ($M$) of the vibration rate, which we will assume here.

To calculate the bandwidth required in order to avoid distortion and possible bias in the estimate of $V_c$, consider a sinusoidal signal with a half-period corresponding to the spacing between the KP signal peaks, $2\varphi_0$. That sinusoid has a frequency $\beta_1 f = \pi f/2\varphi_0$ . We find that by incorporating signals up to a frequency $\beta f = 2\beta_1 f \approx 2\pi f/(\pi - 3\gamma)$, bias is negligible (see Fig 1a).

We now consider a suboptimal estimator. This may be simpler to program and would consume fewer processor cycles, which might be useful with some embedded microprocessors. We use only the most sensitive part of the data and a simplified *ad hoc* model of the form

$$z(\varphi) = \begin{cases} +\hbar(1 + a_2(\varphi + \varphi_0)^2 + a_3(\varphi + \varphi_0)^3 + \ \dots), & \varphi < 0 \\ -\hbar(1 + a_2(\varphi - \varphi_0)^2 - a_3(\varphi - \varphi_0)^3 + \ \dots), & \varphi > 0 \end{cases} \tag{13}$$

The peaks are at $\varphi = \pm\varphi_0$. We use only the data close to the peaks. In particular, using the nominal value of $\varphi_0$, we take all data for which $\varphi_0(1-\kappa) < |\varphi| < \varphi_0(1+\kappa)$, where $0 < \kappa < 1$. Alternatively, we could take all data such that $|F(\gamma,\ \varphi)| > kF_0$ where $0 < k < 1$. Then we estimate h, $\varphi_0$, $a_2$, and additional $a_n$ as needed. Compared to the use of all of the data and the exact model (Eq 10), this analysis will have an increased uncertainty and a bias. A numerical evaluation of these limitations is presented in Section IV.

Whether the real-time estimator is based on Eq 10 or on Eq 13, the estimation problem is non-linear (in $\gamma$ or $\varphi_0$, respectively.) However, the starting value for the non-linear parameter can be the most recent estimate or a smoothed version of that. Thus, we anticipate that the non-linearity will not require iteration of the estimator at every cycle, although iteration will likely be required at the start of the run.

*3.2. Estimation of CPD from separate estimates of $\hbar_1$ and $\hbar_2$ (Stage 2 analysis).*

The above analysis of the phase and height of the peaks can be extended to the calculation of an estimate of the CPD that is suboptimal, but may be adequate for many purposes. We start by finding the average of all data taken with the same phase and backing potential. Then, solving as above separately for parameters associated with each backing potential, we would obtain estimates of $h_1$, and $h_2$ ($\hat{\hbar}_1$ *and* $\hat{\hbar}_2$ ). There would also be a pair of estimates of $\varphi_0$ and of each of the $a_n$ terms, but we discard these.

Starting with the estimates of the $h_i$, we use Eq 4 and 5 to find the CPD.

$$\hat{V}_c = \frac{V_{b1} + V_{b2}}{2} - \frac{\left(\hat{\hbar}_1 + \hat{\hbar}_2\right)(V_{b1} - V_{b2})}{2\left(\hat{\hbar}_1 - \hat{\hbar}_2\right)} \tag{14}$$

If the KP is being operated in the standard mode, the first term on the right is approximately the CPD and the second term is the (small) measured correction. The uncertainty in this estimate of $V_c$ comes from the uncertainties in the estimates of $h_1$ and $h_2$. With ordinary instrumentation, the uncertainty in the backing potentials should make a negligible contribution to $\sigma(V_c)$.

The sensitivity of $V_c$ to $h_i$ is

$$\frac{\partial \hat{V}_c}{\partial \hat{h}_1} = h_2 \frac{V_{b1} - V_{b2}}{(h_1 - h_2)^2}$$
$$\frac{\partial \hat{V}_c}{\partial \hat{h}_2} = -h_1 \frac{V_{b1} - V_{b2}}{(h_1 - h_2)^2} \tag{15}$$

By taking $\hat{h} = \hat{h}_1 = -\hat{h}_2$ and $\sigma(h) = \sigma(h_1) = \sigma(h_2)$, and noting that $\hat{h}_1$ and $\hat{h}_2$ are independently determined and thus uncorrelated we find

$$\sigma(V_c) = \left|\frac{V_{b1} - V_{b2}}{2\, h}\right| \frac{\sigma(h)}{\sqrt{2}} \tag{16}$$

Note that $\hat{h}_1$and $\hat{h}_2$ are each proportional to amplifier gain. By referring to Eq 14, it can be seen that a change of gain does not affect the estimate $\hat{V}_c$. However, should the gain change between the times that $V_{b1}$and $V_{b2}$ were applied, it would result in an error in $\hat{V}_c$.

To find $\sigma(h)$, we revisit the Stage 1 analysis. Starting with a simplified version of Eq 13,

$$z(\varphi) = \begin{cases} h(1 + a_2(\varphi + \varphi_0)^2\,), & \varphi < 0 \\ -h(1 + a_2(\varphi - \varphi_0)^2\,), & \varphi > 0 \end{cases} \tag{17}$$

we perform a covariance analysis with a three element parameter set: {h, $a_2$, $\varphi_0$}. This is conveniently done by replacing the sums over data with the corresponding integrals. The integration covers one vibration cycle (of averaged KP data) in which data are included only over the ranges $\varphi = -\varphi_0 \pm \kappa\, \varphi_0$ and $\varphi = \varphi_0 \pm \kappa\, \varphi_0$. The weighting density (inverse noise power density) takes the form

$$W = \frac{\tau_i}{2\,\pi\,\left(G\,\sigma_j\right)^2} \tag{18}$$

where $\tau_i$ is the time spent at a single backing potential (i = 1, 2) and $\sigma_j$ is the preamplifier input noise. After forming and inverting the normal equations, we find

$$\sigma(h) = \frac{3\,G\,\sigma_j}{4}\,\sqrt{\frac{2\,\pi}{\kappa\,\varphi_0\,\tau_i}} \tag{19}$$

and that the estimate of $\varphi_0$ is uncorrelated with either of the other two estimates. (This results from the selection of data symmetrically around $\varphi_0$.)

We use $h$=$YF_0$ and Eq 10 to show that

$$\mathcal{h} = \frac{\omega\ \epsilon_0 \mathcal{A}\ G\ F_0}{2\ d_0} (V_{b1} - V_{b2}) \tag{20}$$

and, by combining this with Eq 16 and 19, we find the uncertainty for the Stage 2 analysis

$$\sigma(V_c) = \frac{3\ \sigma_j\ d_0}{4\ \omega\ \epsilon_0 \mathcal{A}\ F_0} \sqrt{\frac{2\ \pi}{\kappa\ \varphi_0\ \tau_0}} \tag{21}$$

where we have used $\tau_0/2 = \tau_1 = \tau_2$.

For the measurement uncertainty, we take the measured current noise of a recently built Kelvin probe preamplifier [Phillips et al. in preparation], $\sigma_j = 3\ 10^{-15}$A Hz$^{-1/2}$. (A preamplifier based on this design may soon become available in a commercial instrument.) In evaluating Eq 21, we take $\gamma = 0.75$, which implies $\varphi_0 = 0.458$ and $F_0 = 3.10$. We also take $d_g = 0.01$ mm, which implies $d_0 = 0.04$ mm, to find $\sigma(V_c) = 19.5\ \mu V$. In evaluating Eq 21, it must be remembered that, as $\kappa$ gets large, the estimates of $h_i$ become increasingly biased, although the biases of $\widehat{\mathcal{h}}_1$ and $\widehat{\mathcal{h}}_2$ cancel in the estimate of $V_c$ via Eq 14. The selection of a reasonable value for $\kappa$ is addressed in Section IV.

### *3.3. A better estimate of CPD from simultaneous estimates of $\mathcal{h}_1$ and $\mathcal{h}_2$.*

An improved Stage 1 analysis, combined with the same Stage 2 analysis, gives a better CPD estimate. Above, we used estimates of $h_1$ and $h_2$ found separately. Here we use estimates of $h_1$ and $h_2$ found simultaneously. This leads to a single estimate of $\varphi_0$ and each of the $a_n$ instead of two. The analysis starts with Eq 14 and a covariance analysis with four parameters: $h_1$, $h_2$, $\varphi_0$ and $a_2$. From that covariance matrix, we take $P_2$, the 2 x 2 sub-matrix for the parameters $h_1$, and $h_2$. Because $\hat{V}_c$ does not depend on the other two parameters estimated, we can extract $P_2$ and transform it to the variance in $V_c$, using $J_2$, the Jacobian

$$J_2 = \begin{pmatrix} \frac{\partial \hat{V}_c}{\partial \hat{h}_1} \\ \frac{\partial \hat{V}_c}{\partial \hat{h}_2} \end{pmatrix} \tag{22}$$

Then, the variance of $V_c$ is

$$\left(\sigma(V_c)\right)^2 = var(V_c) = \tilde{P} = J_2^{\dagger} P_2\, J_2 \tag{23}$$

The corresponding standard deviation is

$$\sigma(V_c) = 13.\, \mu V\ \frac{\sigma_j}{3\ 10^{-15}\ A/\sqrt{Hz}}\ \frac{d_0}{0.04\ mm}\ \frac{100\ \ s^{-1}}{\omega/(2\ \pi)} \tag{24}$$

$$\times \frac{\pi\, mm^2}{\mathcal{A}} \frac{3.10}{F_0} \sqrt{\frac{0.1}{\kappa}} \sqrt{\frac{0.458}{\varphi_0}} \sqrt{\frac{1\, s}{\tau_0}}$$

We will use this in Section V to compare to the numerical results from the simulations. The improvement by a factor of 1.5 comes from the correlation of $h_{1,}$ and $h_2$ in $P_2$ (ρ = -0.385), and from a smaller variance of $h_{1,}$ and $h_2$ by a factor of 13/18 = 0.722.

## 4. Simulation of the Real Time Estimator

The principal function of the real-time (Stage 1) estimator is to provide $\hat{\varphi}_0$ to a servo that maintains $d_0$. A second possible use is to observe rapidly changing values of CPD. We have investigated the properties of the real-time estimator via a series of simulations. In all of the simulations, we assumed that the input amplifier had a white input current noise with density 3 $10^{-15}$ Amp/Rt(Hz), and that the signal is sampled at sufficiently short intervals that the results are free of sampling artefacts. We further assumed a cylindrical tip of radius 1 mm with a flat end perpendicular to the cylinder axis and parallel to the sample surface, a tip vibration along the cylinder axis at a frequency of 100 Hz and a fixed gap (minimum tip-to-sample distance, $d_g = d_0 - d_1$ of 0.01 mm. The mean spacing is then given by $d_0 = d_g/(1-\gamma)$. We have used the simulation to estimate both the standard deviation and the bias in the estimate of both $\varphi_0$ and $h$ based on data near the (positive and negative) peaks of 10 cycles of tip vibration (0.1 s) such that $\varphi_0(1-\kappa\,) < \varphi < \varphi_0(1+\kappa)$ or $-\varphi_0(1+\kappa) < \varphi < -\varphi_0(1-\kappa)$ as discussed above.

In performing the first simulations of Stage 1, pseudo data were found using Eq. 10 and the fitting model was given by Eq 17, with parameters h, $\varphi_0$ and $a_2$. Figures 3 and 4 show the normalized uncertainty in $\hat{\varphi}_0$ and $\hat{h}$, respectively, as a function of $\gamma$ and for several values of $\kappa$. As expected, letting the tip come closer to the sample (making $\gamma$ larger and holding $d_0$ constant) increases the precision of the estimate of either $\varphi_0$ or $h$. Over the range of $\gamma$ shown, $\sigma(\varphi_0)/\varphi_0$ is dropping almost linearly with increasing $\gamma$. The strong dependence of $\sigma(\varphi_0)/\varphi_0$ on κ is related to the increased sensitivity $(\delta i/\delta\varphi_0)$ when $\varphi$ is further from $\varphi_0$. One expects that $\sigma(\varphi_0\,)/\varphi_0 \;\propto \kappa^{-3/2}$ since the sensitivity of the data increase linearly with κ. This is confirmed by the simulations. The pattern is similar for $\sigma(\hat{h})/\hat{h}$ except that there is no increase in sensitivity as $\kappa$ increases, just an increase in total data i.e., $\sigma(\hat{h})/\hat{h} \propto \kappa^{-1/2}$. This also is confirmed by the simulations.

Figures 5 and 6 are similar to Figs 3 and 4, but show the bias in the estimate. The biases can be seen to be nearly independent of $\gamma$ but to depend strongly on the width of the data window, which is not surprising. The value of $h$, and thus of $\hat{h}$, scales with $V_b - V_c$ (Eq 5). As long as the shape of the signal waveform $F(\gamma, \varphi)$ is unchanged, the bias in $\hat{h}$ will be proportional to h.

$$< \hat{h} >= (1+\epsilon)h \quad (25)$$

where $\epsilon$ is the quantity shown in Fig. 6 and angle brackets denote expectation value. By applying Eq 25 to Eq 14, we see that the estimate of $V_c$ is independent of the bias in $\hat{h}$. The bias in $\hat{\varphi}_0$ does not affect the estimate of $V_c$ to first order, but will affect the tip sample spacing if it is used in the distance servo. As long as the bias in $\hat{\varphi}_0$ does not change, it will not matter in the servo application, nor will it contribute uncertainty to $\hat{V}_c$.

Figures 7, 8, 9, and 10 parallel the previous four, but show the interaction of the value taken for $\kappa$ with the number of terms in the time series, Eq 13. For both $\sigma(\varphi_0)/\varphi_0$ and $\sigma(h)/h$ we find that the addition of a cubic term has no effect, likely because of the symmetric way in which the data are selected around $\varphi_0$. However, the addition of a quartic term to the estimator results in about a 2.5 fold increase in $\sigma(\varphi_0)/\varphi_0$ and a much smaller increase (about 20%) in $\sigma(h)/h$. The effect on bias of adding the cubic and quartic terms is far greater. In the case of bias, it is the cubic term that makes the difference, with the addition of the quartic term having no further effect. The effect of the cubic term is largest for small values of $\kappa$. With increasing data window, the bias increases and the reduction of bias by adding terms to the model decreases.

For fixed $d_g$, increasing $\gamma$ from 0.3 to 0.9, improves $\sigma(h)$ and therefore $\sigma(V_c)$ only modestly, about three fold. Over a more restrictive range ($0.7 < \gamma < 0.9$) in which one could plan to operate, $\sigma(V_c)$ varies by $\pm$ 20%, and thus the selection of $\gamma$ is not likely to be critical. The variation of $\sigma(\varphi_0)$ with $\gamma$ is even smaller. The issue of bias applies only to the approximate models discussed above for use in a computation-limited real-time environment.

## 5. Suboptimal Estimator for Scientific Results

The best estimates of CPD and the mean spacing of the vibrating capacitor ($d_0$) will come from fitting the parameters of Eq 10 ($V_c$, $\gamma$ and $d_0$) using the data taken with two backing potentials, preferably following the standard KP operating mode described in Section II. However, we find from Eq 24 that, for a 1 s observation, the precision with the suboptimal estimator can be ~13 $\mu$V. For many purposes, this is a far higher resolution than is needed. Here we consider using the real-time data analysis, developed to stabilize the mean tip spacing, to determine $h_1$ and $h_2$. We then apply Eq. 14 to find a good but not optimal estimate of CPD.

The algorithm we investigated starts with the accumulation of the data used for the real-time analysis. We assume that the data sampling is synchronous with the vibration of the tip, which makes the accumulation easy. The data taken with both backing potentials are used to estimate $h_1$, $h_2$, $\varphi_0$, $a_2$, and any higher $a_n$ needed. This is a non-linear problem and thus may require iteration, but the parameter estimates found in real time provide good starting values for the estimator. As a practical matter, it may be useful to estimate an additional parameter representing an overall phase shift between the synchronization signal and the digitized KP signal. Such a phase shift could come from the analogue electronics or the mechanical tip-vibration system. By symmetry, when this parameter is estimated, it should be uncorrelated with the others. Thus, compared to making the phase correct by external adjustment, adding the parameter should have no effect on either the other parameter estimates or their uncertainties. The resulting estimates of the $h_i$ are used along with the known $V_{bi}$ to estimate $V_c$ according to Eq 14. The standard deviation of $\hat{V}_c$ is found by propagation of error from the covariance matrix resulting from the above estimation.

Figure 11 shows the uncertainty in $\hat{V}_c$ as a function of $\gamma$ and for two values of $\kappa$. Also shown for comparison is the optimal estimate based on fitting the parameters of Eq 10 to all of the data $(-\pi < \varphi < \pi)$. In all cases, changing the modulation index $\gamma$ from 0.3 to 0.9 produces about a factor of 3 reduction in $\sigma(V_c)$. Based on the results for $h$, we had expected $\sigma(V_c)$ to depend on the number of $a_n$ terms included in the analysis. However, we find that $\sigma(V_c)$ is independent of whether we cut off the polynomial series at $a_2$, $a_3$, or $a_4$ (to within a few parts in $10^6$, and this is non-zero only because of limited numerical accuracy. See next section.) Further, in the case $\gamma = 0.75$, we compared the numerical result (top line, Fig 11) with an analytic determination of $\sigma(V_c)$ from Eq 24 and found agreement to about $10^{-3}$. Two factors interact to produce the constancy of $(V_c)$ as the number of $a_n$ terms is changed, $\sigma(\hbar_1) = \sigma(\hbar_2)$ and $\rho(\hbar_1,\ \hbar_2)$, both of which change with a change of parameter set. Referring to Fig 11, we note that the optimal estimator is better than the cases of $\kappa = 0.1$ and $\kappa = 0.5$, but only by a factor of ≈2.5 and ≈1.3, respectively.

## 6. Discussion

We have described a novel approach to the stabilization of $d_0$, the mean spacing of the KP tip from the measured surface. A more traditional approach is based on the Fourier decomposition of the signal from the preamplifier and the use of the ratio of successive terms as a measure of $d_0$. As $\gamma$ approaches 1, the signal is spread over an increasingly large number of Fourier terms, which suggests that large $\gamma$ may be inconsistent with good stabilization by the Fourier ratio method.[1] In either case, the adequacy of the stabilization will depend on the disturbance to $d_0$, which is beyond the scope of the present paper.

[1] In the case of large $\gamma$, there would be an advantage to estimating $\gamma$ from terms $a_i$ and $a_{i+n}$, where $i$ and $n$ are chosen to encompass the large terms among the $a_j$. Further, that estimate could be combined with the estimate from terms $a_{i+1}$ and $a_{i+n-1}$, etc. Additional analysis is required.

In the previous section, we found surprisingly from numerical analysis that $\sigma(V_c)$ is independent of the degree of polynomial (terms $a_2$, $a_3$, $a_4$) included in the model. In order to shed light on this result, we developed an analytic covariance matrix based on a data schedule that included two backing potentials equally spaced above and below $V_c$ with equal time at each $V_b$. The parameter set is $\{\hbar_1, \hbar_2, \psi\}$, where $\psi = \{\varphi_0, a_j, j = 2, \ldots, n\}$. $\psi$ characterizes the shape of the peaks of the signal model, and $n$ is the highest order polynomial term included. By symmetry, $\partial z/\partial \hbar_1 = -\partial z/\partial \hbar_2$, and this will be reflected in the elements of the $(2+n) \times (2+n)$ covariance matrix. (In the notation used below, the subscript is the size of the matrix for the case $n = 2$.)

$$P_4(\hbar_1, \psi) = -P_4(\hbar_2, \psi). \tag{26}$$

We can construct a Jacobian to reduce the parameter count by one, $\{\hbar_1, \hbar_2, \psi\} \rightarrow \{V_c, \psi\}$. For example, for n=2,

$$J_4 = \begin{pmatrix} \eta_1 & 0 & 0 \\ \eta_2 & 0 & 0 \\ 0 & 1 & 0 \\ 0 & 0 & 1 \end{pmatrix} \tag{27}$$

where $\eta_i = (\partial V_c)/\partial \hbar_i$ $(i = 1,2)$ and it is easily shown that $\eta_1 = \eta_2$. Proceeding as in Eq 23, we obtain the $(1+n) \times (1+n)$ covariance matrix, for the new parameter set, for example

$$\tilde{P}_3 = J_4^{\dagger} P_4 J_4. \tag{28}$$

We find that all of the covariance terms of the form $\tilde{P}_3(V_c, \psi)$ are zero. Since $V_c$ is uncorrelated with any of the other parameters, $\sigma(V_c)$ is independent of the number of terms in the polynomial series.

It was noted in Section IV that $\hat{V}_c$ (Eq 14) is unbiased, and in particular, that it is independent of the bias in $\hat{\hbar}$. The terms $a_3$ and $a_4$ were originally included in the analysis to reduce the bias in the estimate of $h$ in the expectation that this would decrease the bias in $\hat{V}_c$. We must now ask whether it matters how many $a_n$ terms (n > 2) are included in the estimate of $h$. There are three reasons for setting n > 2 when $\kappa$ has not been set very small. A model error increases the actual variance (cf. the root variance estimate from the WLS inverse coefficient matrix) of the estimates of $V_c$ and $\varphi_0$. If the value of $\varphi_0$ estimated from the data is to be used for other than stabilizing $d_0$, then the terms with n > 2 should be used to limit the bias in $\hat{\varphi}_0$ as needed. A standard method of determining or verifying the level of measurement noise is to examine the post-fit residuals. If the model fails to fit the data, the post-fit residuals will be excessively large. These three reasons notwithstanding, in many of the uses of the KP, particularly when its highest accuracy is not needed, the effects of model error will not matter and operating with n = 2 will suffice.

Popescu [2011] investigated a KP analysis scheme based on WLS fitting to a simplified model. That paper describes the peak (and anti-peak) of the signal with the equivalent of our parameter $a_2$, but not higher-order polynomial terms. He introduced a data acceptance-zone parameter, *f*, (which we called *k*, see description below Eq 13) such that data, $z(\varphi)$, would be included only when $|z(\varphi)| > f\ z_{max}$. He found large systematic error in the estimate of $h$ over the range of *f* that he considered, $f \leq 0.8$ (see Fig 1). This corresponds approximately to $\kappa > 0.45$ (since $\kappa \approx \sqrt{1-f}$ ). We also find large biases for the case that $\kappa = 0.5$ and no polynomial term beyond $a_2$ is included. As noted above, for $\kappa = 0.1$, we find a modest bias. Popescu also finds a bias in the estimate of $V_c$ (his Fig 16). This is in contrast with our conclusion that $\hat{V}_c$ is unbiased when found using Eq 14.

It has been claimed in the past that it is not practical to make KP measurements with $V_b = V_c$, because in this condition, "the signal to noise ratio is zero." This is an incorrect application of the concept of an SNR. Some of the most precise experiments ever conducted are "null tests" in which there is no need to subtract a calculated quantity from that measured in order to detect or bound an anomaly. That the claim is false can be seen from Eq 9. In the standard KP operating mode defined here, it is necessary to have $V_b \neq V_c$ in order to derive a signal for the servo that holds the tip-sample distance constant. Therefore, to work with $V_b = V_c$, we need independent distance information, but we can get it by applying a separate signal to the tip at a frequency very different from (much higher than) that of the tip oscillation, or by periodically interrupting observations used for determining $V_c$ and applying $V_b \neq V_c$ in order to determine distance.

A graph has even been published in support of the notion that we must have $V_b \neq V_c$, showing that as $|V_b - V_c|$ is reduced, the KP signal, which would be expected to go to zero in proportion to $|V_b - V_c|$, instead behaves erratically. Eq. 9 shows that the signal does not behave erratically in a properly designed apparatus. However, it might do so if the measurement apparatus was unable to determine the phase of the KP signal *ab initio*¸ and had to do so instead *from the signal.* The measurement apparatus does not need to determine the phase from the signal: at fixed tip oscillation frequency, the required phase has a fixed offset from the phase of the tip drive.

It has also been claimed that taking data with more than two values of $V_b$ is helpful in measuring the slope of the line $i$ vs. $V_b - V_c$. This is also false. It was demonstrated in Sec. II that the sensitivity to $V_c$ is not increased by using more than two points. There is a small cost to using more than two points: the time required for the input stage to settle after a change in $V_b$.

## 7. Conclusions and future directions

A common problem for gravity experiments that detect extremely small low-frequency or constant forces is the presence of spurious non-gravitational forces that can neither be calculated nor separately measured. In SR-POEM, as well as in LISA and LISA Pathfinder (Antonucci et al. 2012), patch effect forces must be suppressed by fabricating surfaces that have small spatial variation of potential. For SR-POEM, we also require that the surface potential be stable in time. In order to be able to fabricate such surfaces, it is essential that the spatial variation of potential be measurable and that the required measurement times be short. Such measurements would be made with a KP. Up to now, the limit to KP measurements had been about 1 mV, and such measurements could not be made in 1 s as required for SR-POEM. SR-POEM will be conducted in a vacuum of $10^{-10}\ Torr$ in order to suppress the noise due to squeeze-film damping. This will limit the change of potential due to surface contamination.

We have investigated aspects of the processing of the digitized KP data, with emphasis on the effect on the uncertainty in the measured CPD of the user's choices of hardware and analysis parameters. We have shown that there is no need to have more than two backing potentials for a given spot on the measured surface, and that if the measurement times are matched to the backing potentials, the uncertainty in $\hat{V}_c$ depends on the two measurement times only through their sum. In order to maximize the precision of $\hat{V}_c$, the peak signal heights should be estimated simultaneously from the two sets of digitized signals taken with the two backing potentials. Independently calculated estimates of the $h_i$ yield a $\sigma(V_c)$ that is larger by a factor of 1.5. Further, we have shown that, when the $h_i$ are simultaneously estimated, $\hat{V}_c$ is not correlated with the $\hat{a}_n$, which removes one of the barriers to including a sufficient number of these terms, since they will not degrade $\hat{V}_c$. Including a sufficient number of $\hat{a}_n$ reduces the bias in $\hat{\varphi}_0$. As we have shown, $\hat{V}_c$ is unbiased.

A further advance in KP sensitivity might come from the introduction of a Kalman filter processing of the real-time data and the corresponding extraction of a measure of the tip spacing. This would allow a closer spacing and a correspondingly larger signal without increasing measurement noise. This approach would put a more stringent requirement on the deviation of the tip from being parallel to the measurement surface. Finally, a major advance in the KP sensitivity will come from an improved preamplifier. A description of such a preamplifier will be published elsewhere by Phillips and Reasenberg.

## Acknowledgements

This work was supported in part by the NASA Science Mission Directorate through grant NNX08AO04G. Additional support came from the Smithsonian Institution and McAllister Technical Services. We thank E.M. Popescu for his helpful comments on the paper. The simulation code was written in Mathematica® 8.

## Figures and Captions

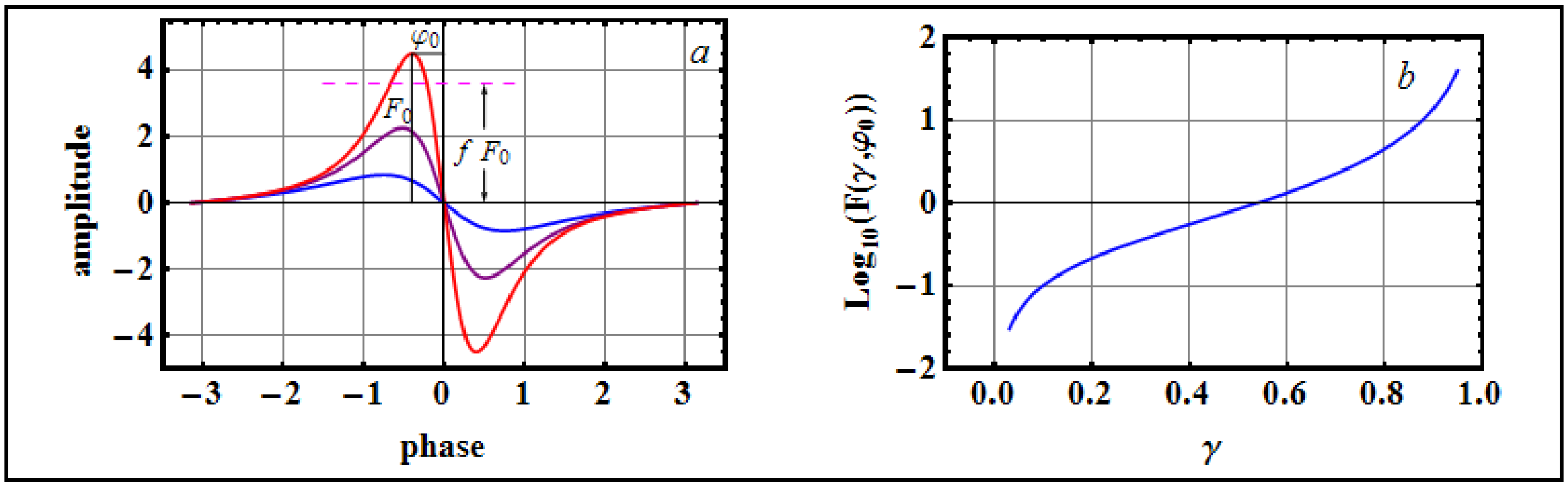

Figure 1. (a) $F(\gamma, \varphi)$ shown over a full cycle for $\gamma$ = 0.5, 0.7, and 0.9 (blue, purple, red). Larger values of $\gamma$ result in higher and sharper extreme heights. Also, for $\gamma = 0.9$, the largest value ($f$ = 0.8) Popescu considered for his cutoff ($f$, which we call $k$) is shown in magenta (dashed). (b) Log ($F_0$).

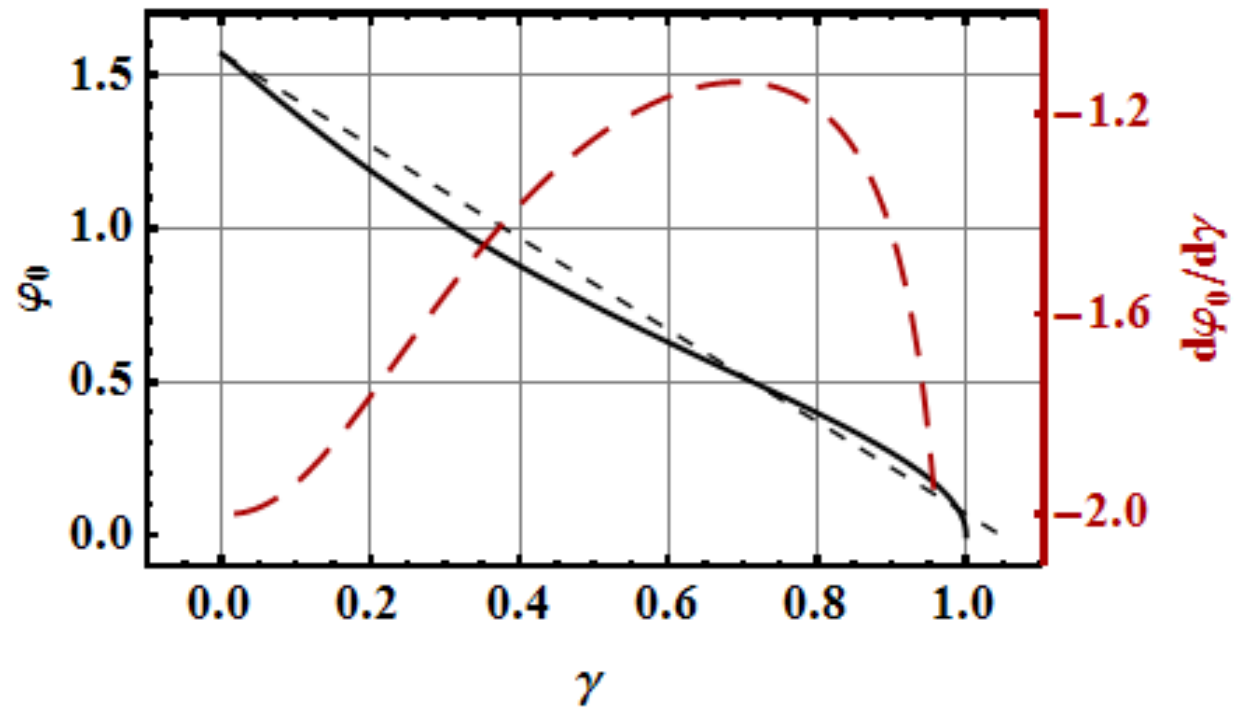

Figure 2. $\varphi_0(\gamma)$ (solid, black, left scale), its linear approximation (thin straight line, short dash, black) and $d\varphi_0/d\gamma$ (long dash, red, right scale).

For the figures below, we will provide versions without the figure number and title at the top. These will be suitable for publication.

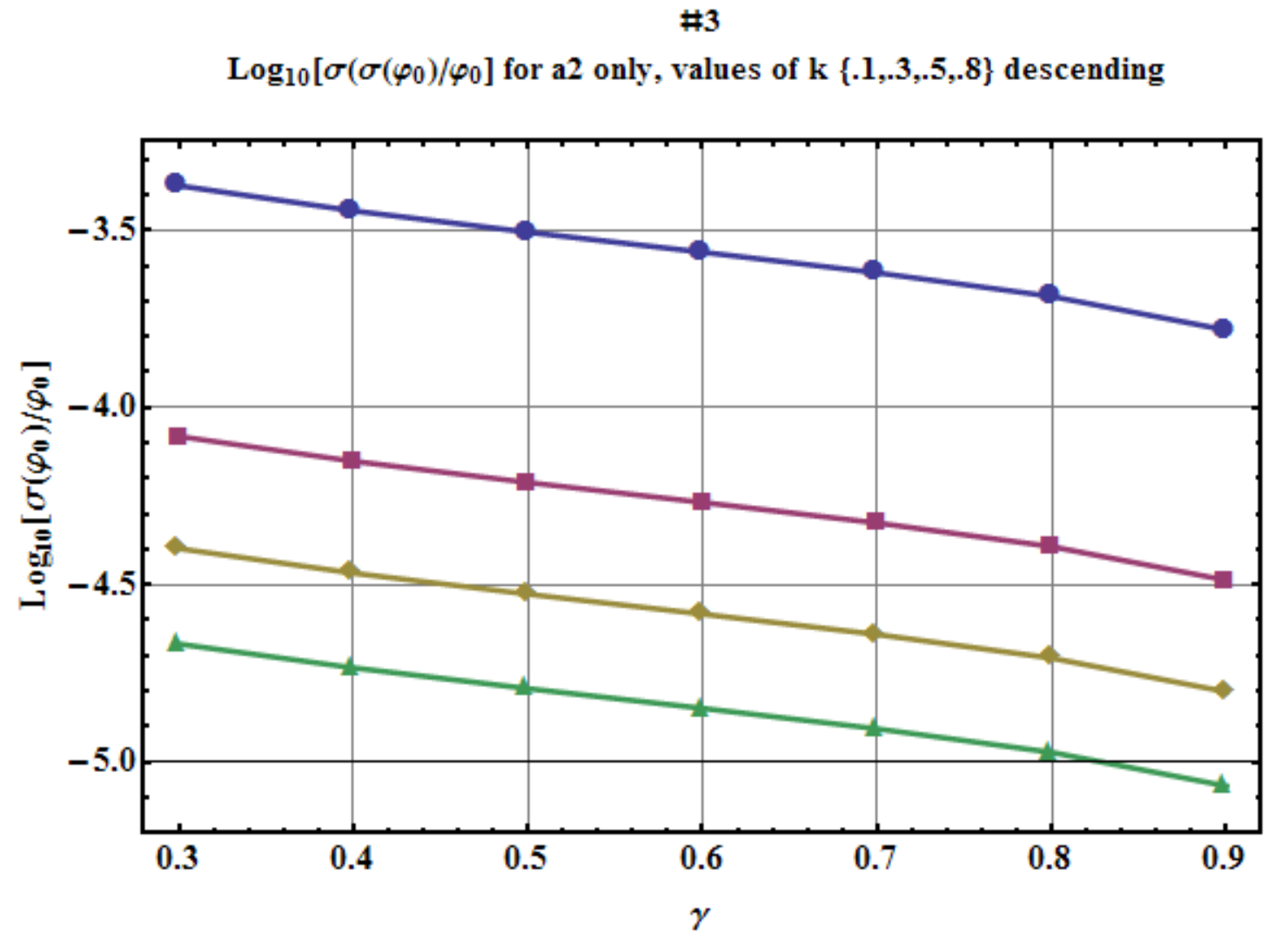

Figure 4. Normalized uncertainty in the estimate of $\varphi_0$ from an analysis based on Eq 17 and shown for four values of $\kappa$, {0.1, blue circle; 0.3, purple square; 0.5, orange diamond; 0.8, green triangle}. As $\gamma$ varues, $d_0$ is adjusted to hold $d_g = d_0 - d_1$ constant. The uncertainty decreases monotonically with $\kappa$.

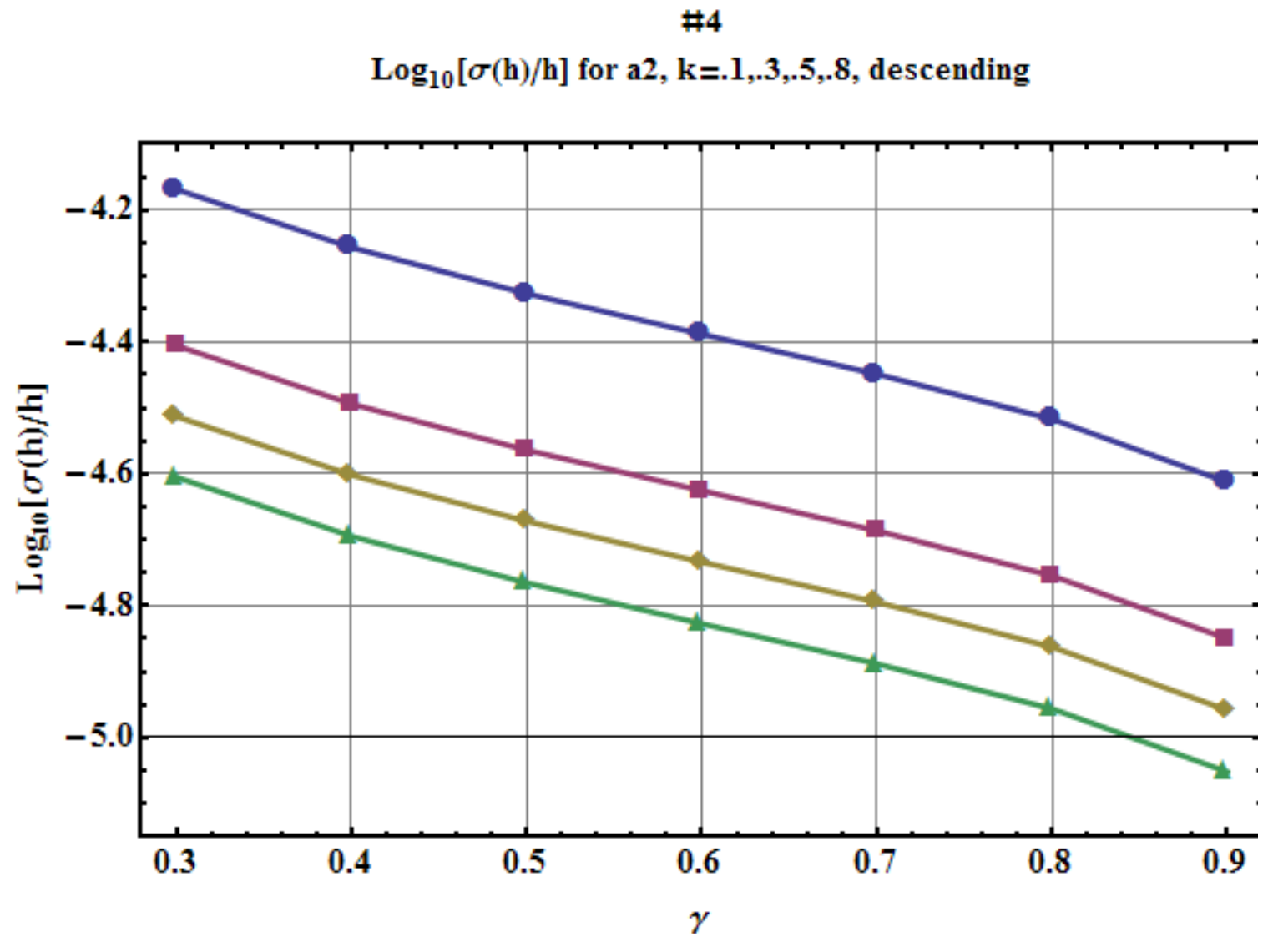

Figure 3. Normalized uncertainty in the estimate of $h$ from an analysis based on Eq 17 and shown for four values of $\kappa$, {0.1, blue circle; 0.3, purple square; 0.5, orange diamond; 0.8, green triangle}. The uncertainty decreases monotonically with $\kappa$.

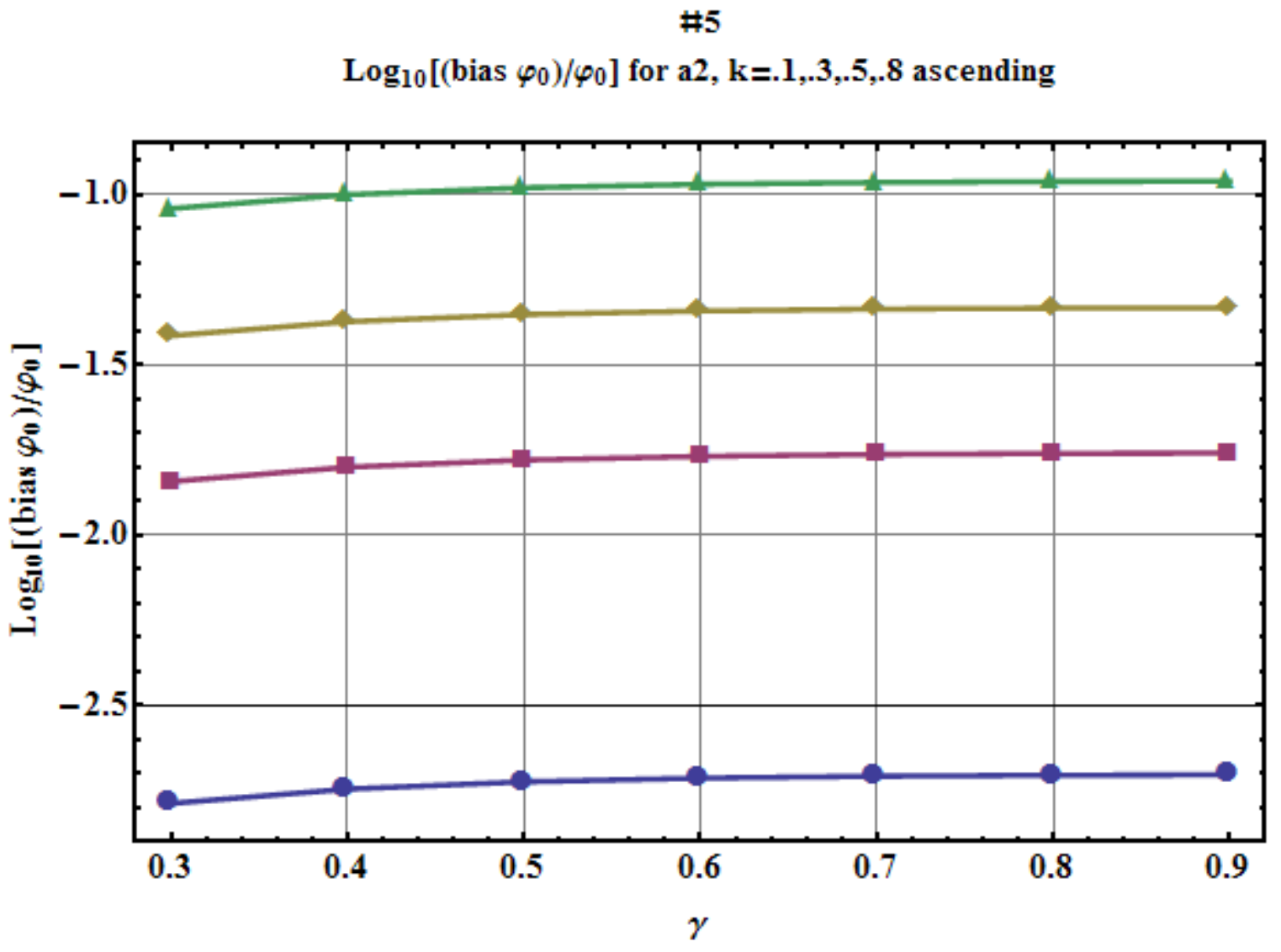

Figure 5. Normalized bias in the estimate of $\varphi_0$ from an analysis based on Eq 17 and shown for four values of $\kappa$, {0.1, blue circle; 0.3, purple square; 0.5, orange diamond; 0.8, green triangle}. The bias is monotonic in $\kappa$ .

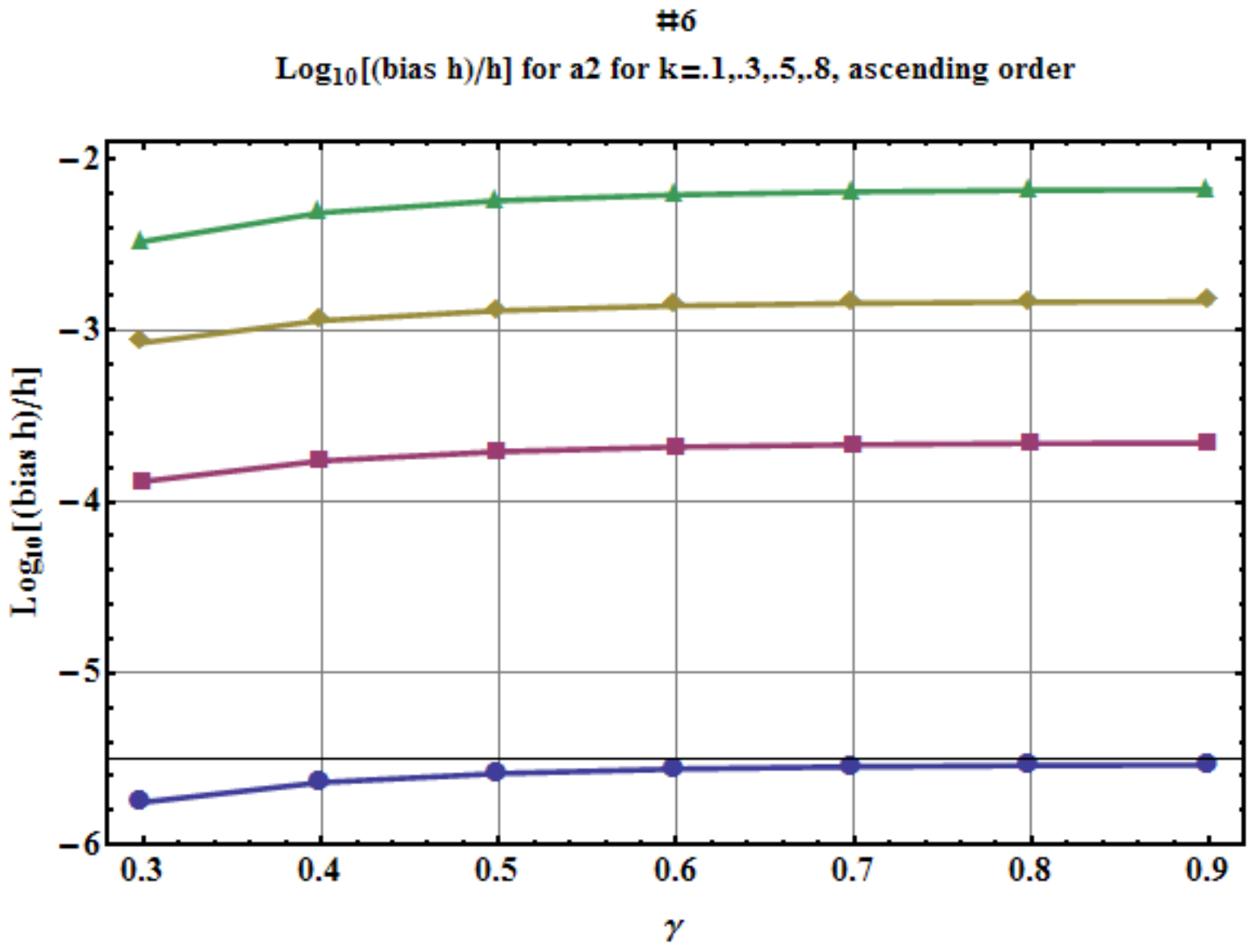

Figure 6. Normalized bias in the estimate of $h$ from an analysis based on Eq 17 and shown for four values of $\kappa$, {0.1, blue circle; 0.3, purple square; 0.5, orange diamond; 0.8, green triangle}. The bias is monotonic in $\kappa$.

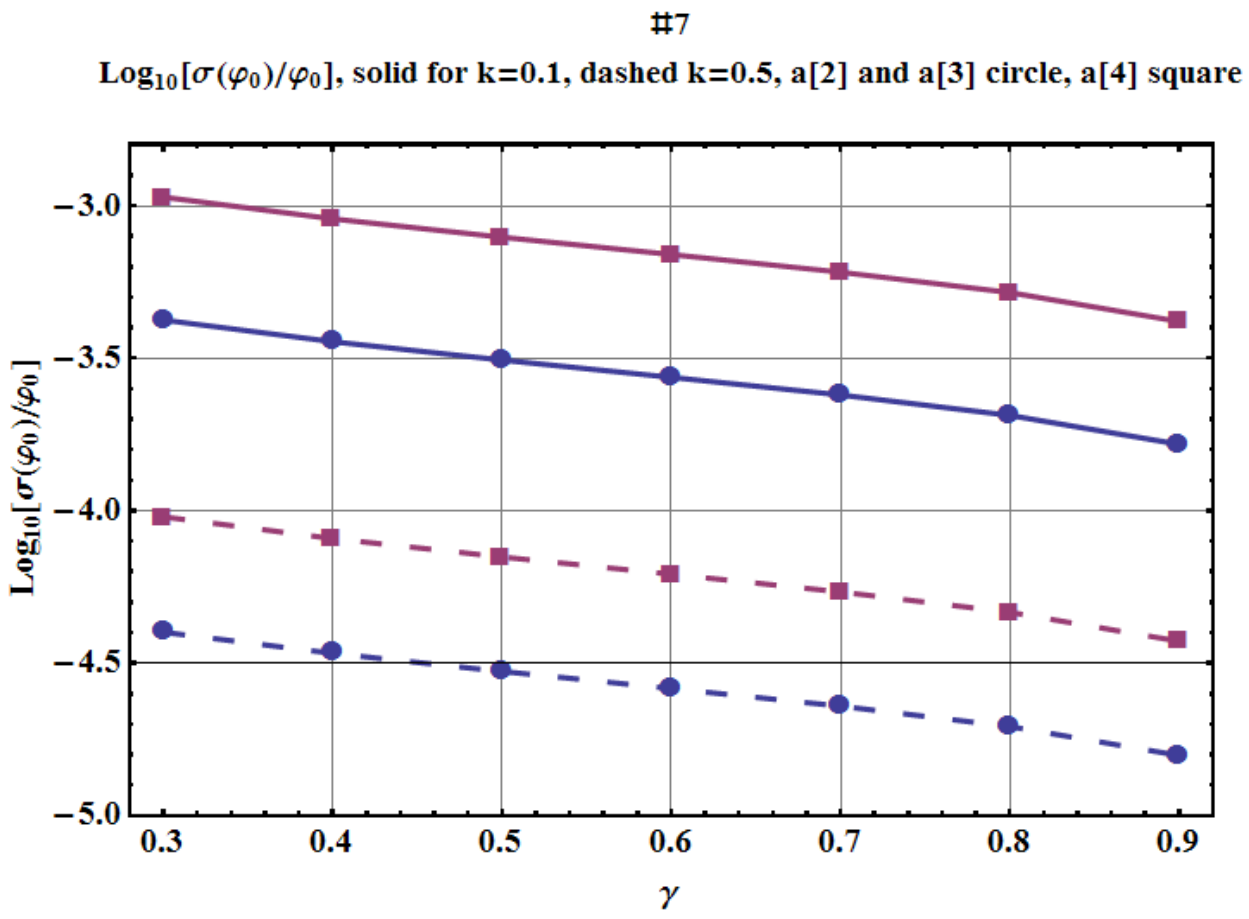

Figure 7. Normalized uncertainty in the estimate of $\varphi_0$ from an analysis based on Eq 13 and shown for two values of $\kappa$: solid lines for $\kappa$ = 0.1 and dashed lines for $\kappa$ = 0.5. The uncertainty decreases monotonically with $\kappa$. Lower curves (blue circle) have quadratic term only in the estimator. Adding the cubic term makes no change. Upper curves (purple square) have quartic term additionally.

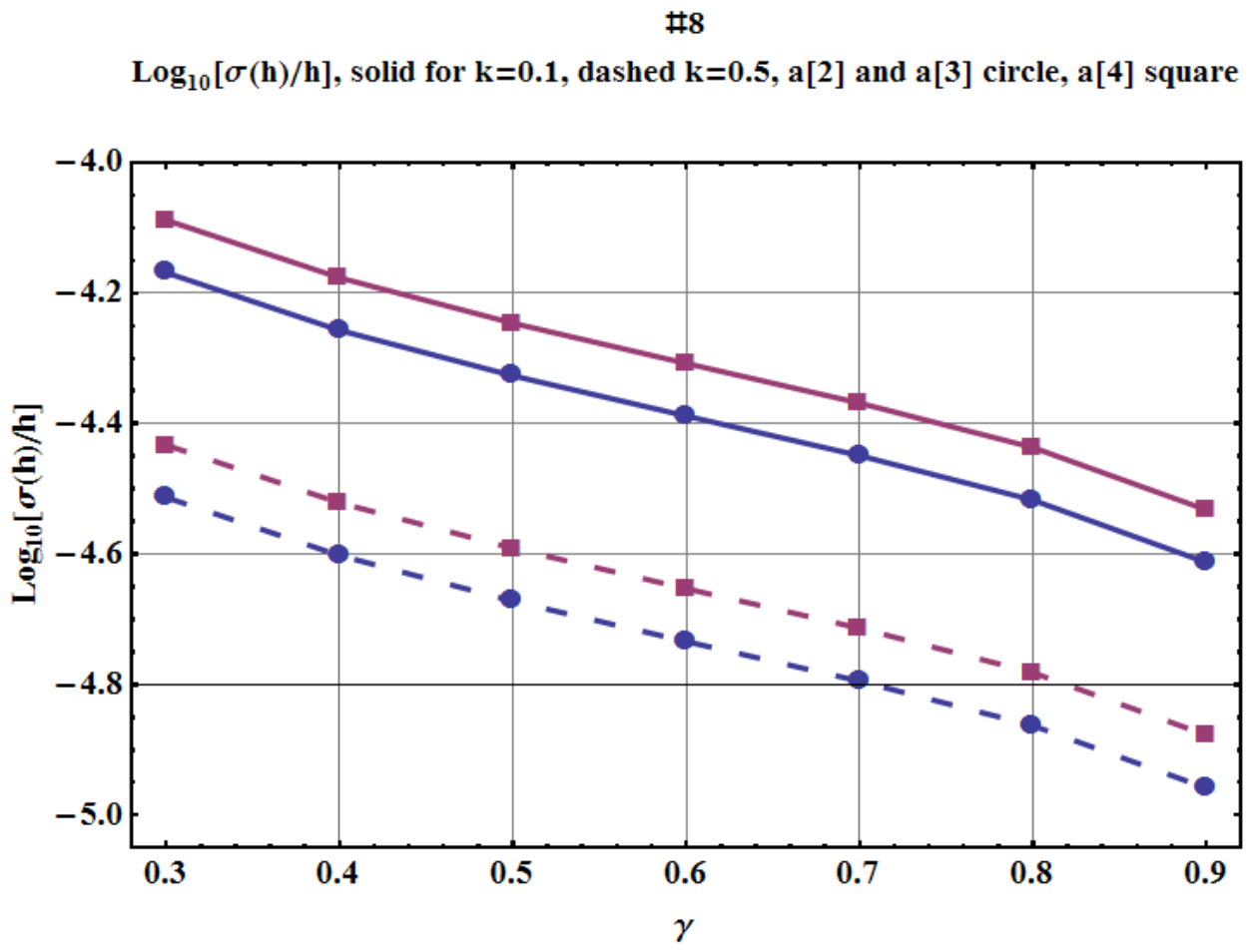

Figure 8. Normalized uncertainty in the estimate of $h$ from an analysis based on Eq 13 and shown for two values of $\kappa$: solid lines for $\kappa$ = 0.1 and dashed lines for $\kappa$ = 0.5. The uncertainty decreases monotonically with $\kappa$. Lower curves (blue circle) have quadratic term only in the estimator. Adding the cubic term makes no change. Upper curves (purple square) have quartic term additionally.

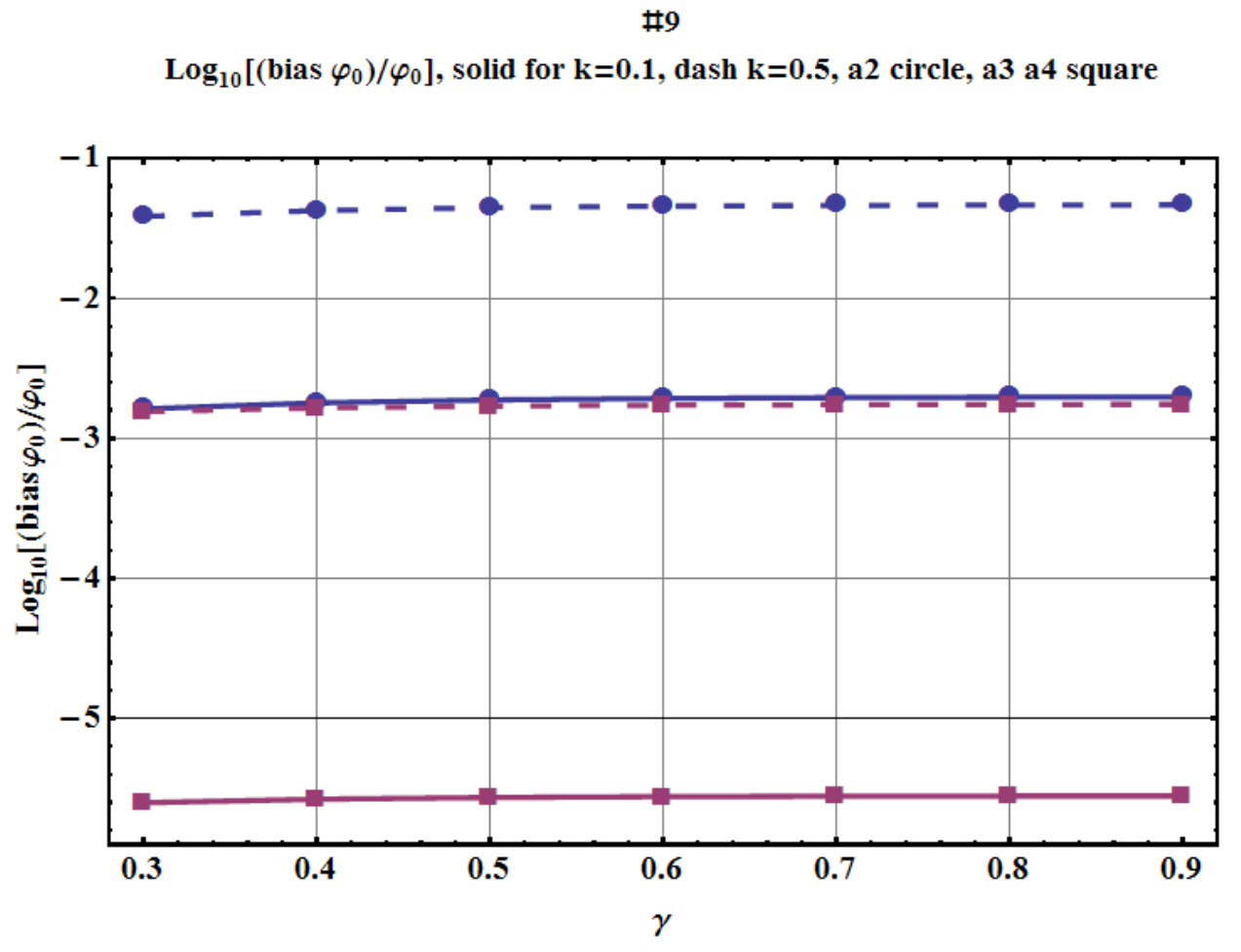

Figure 9. Normalized bias in the estimate of $\varphi_0$ from an analysis based on Eq 13 and shown for two values of $k$: solid lines for $\kappa$ = 0.1 and dashed lines for $\kappa$ = 0.5. The bias increases monotonically with $\kappa$. Upper curves (blue circle) have quadratic term only in the estimator. Lower curves (purple square) have cubic term additionally. Adding the quartic term makes no change. (The middle two curves nearly overlap: $\kappa$ = 0.1, model through $a_2$ and $\kappa$ = .5, model through $a_3$.)

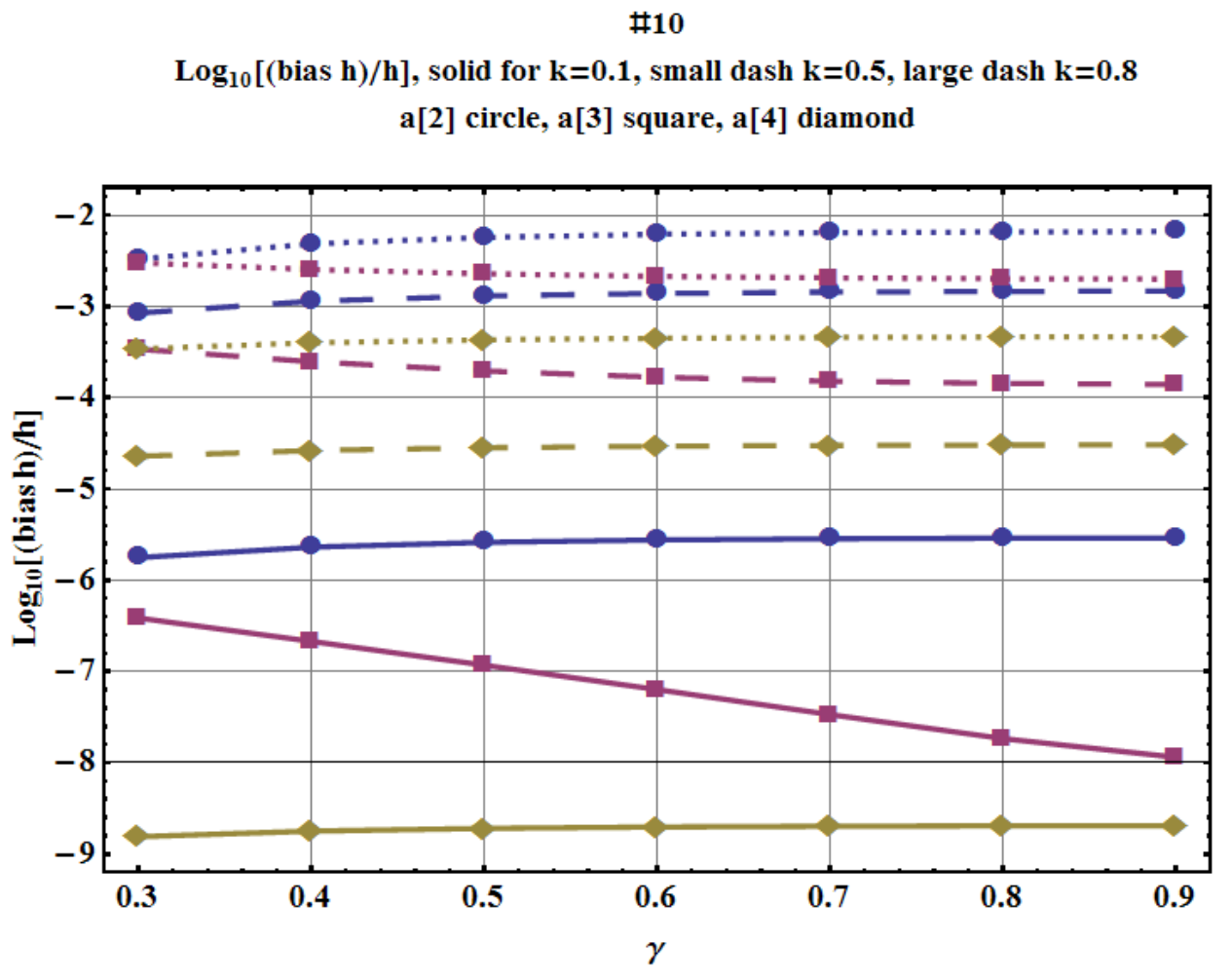

Figure 10. Normalized bias in the estimate of $h$ from an analysis based on Eq 13 and shown for three values of $\kappa$: solid lines for $\kappa$ = 0.1, long dashed for $\kappa$ = 0.5 and short dashed lines for $\kappa$ =0.8. Bias decreases with the addition of cubic and quartic terms to the estimation model. Only $a_2$, blue circle; add $a_3$, purple square; add $a_4$, green diamond. Bias increases monotonically with $\kappa$

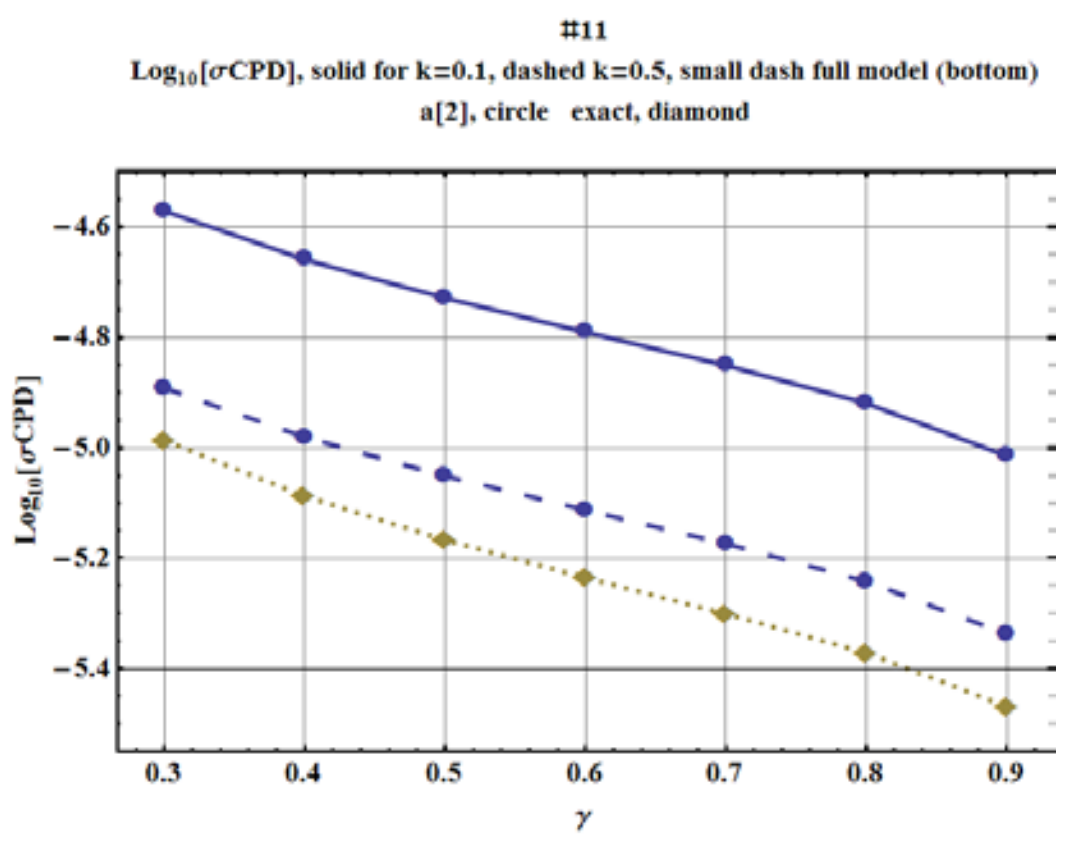

**Figure 11.** Standard deviation of the estimate of CPD for two values of $\kappa$ {0.1, solid; 0.5, dashed} (blue circles) assuming the model contains $a_2$ only, and for the full model (dotted, orange diamonds).